# Quantifying the Value of Real-time Geodetic Constraints on Earthquake Early Warning using a Global Seismic and Geodetic Dataset

**C. J. Ruhl [1,2], D. Melgar [3], A. I. Chung [1], R. Grapenthin [4], and R. M. Allen [1]**

[1] University of California, Berkeley, Berkeley, CA, USA
[2] The University of Tulsa, Tulsa, OK, USA
[3] University of Oregon, Eugene, OR, USA
[4] New Mexico Institute of Mining and Technology, Socorro, NM, USA

contact: cruhl@utulsa.edu

**Key Points:**

- Comprehensive, joint seismic and geodetic analysis of 32 M>=6 global earthquakes to test earthquake early warning (EEW) performance
- Seismically-triggered geodetic algorithms provide more accurate source estimates and ground motion predictions than seismic-only EEW systems
- A coupled system predicts shaking more accurately on a per-station basis and provides higher cost savings performance

**Abstract**

Geodetic earthquake early warning (EEW) algorithms complement point-source seismic systems by estimating fault-finiteness and unsaturated moment magnitude for the largest, most damaging earthquakes. Because such earthquakes are rare, it has been difficult to demonstrate that geodetic warnings improve ground motion estimation significantly. Here, we quantify and compare timeliness and accuracy of magnitude and ground motion estimates in simulated real time from seismic and geodetic observations for a suite of globally-distributed, large earthquakes. Magnitude solutions saturate for the seismic EEW algorithm (we use ElarmS) while the ElarmS-triggered Geodetic Alarm System (G-larmS) reduces the error even for its first solutions. Shaking intensity (MMI) time series calculated for each station and each event are assessed based on MMI-threshold crossings, allowing us to accurately characterize warning times per-station. We classify alerts and find that MMI 4 thresholds result in only 12.3% true positive (TP) alerts with a median warning time of $16.3 \pm 20.9$ s for ElarmS, but 44.4% TP alerts with a longer median warning time of $50.2 \pm 49.8$ s for G-larmS. The geodetic EEW system reduces the number of missed alerts for thresholds of MMI 3 and 4 by over 30%. If G-larmS was triggered instantaneously at the earthquake origin time, the performance statistics are similar, with slightly longer warning times and slightly more accurate magnitudes. By quantifying increased accuracy in magnitude, ground motion estimation, and alert timeliness; we demonstrate that geodetic algorithms add significant value, including better cost savings performance, to EEW systems.

## 1 Introduction

The concept underpinning earthquake early warning (EEW) is to detect and characterize earthquakes as soon as possible after they initiate in order to warn ahead of the arrival of strong ground shaking (*Allen et al.,* 2009a). Ideally, EEW should be a ground-motion (GM) warning system as it is knowledge of the expected intensity of shaking that is most important to a user. In fact, that user's actions often depend on the level of ground motion expected at their site. The United States' ShakeAlert EEW system currently uses earthquake source parameters in combination with a GM model to provide warnings. Therefore, the success of the EEW system depends on accurate earthquake source characteristics (origin time, location, magnitude, and fault-finiteness) in order for the GM estimates themselves to be accurate. Here, we use data from 32 large (M>6) globally distributed earthquakes to evaluate the accuracy and timeliness of earthquake magnitude and ground motion estimates when comparing seismic and geodetic EEW systems. We find that for large earthquakes the performance is much improved when including information from geodetic algorithms.

Traditionally, EEW systems use features of elastic waves recorded on inertial seismometers to estimate the magnitude and epicenter of an earthquake (*Allen et al.,* 2009a). The ShakeAlert system's Earthquake Point-source Integrated Code (EPIC), for example, uses the amplitude of the first few seconds of the P wave arrival on several seismic stations to estimate source parameters (*Chung et al., 2017*). It has been noted that this approach leads to saturation, an underestimation of the magnitudes of large events (*e.g., Hoshiba & Ozaki*, 2014). This is due to saturation of accelerations at higher frequencies in the epicentral region of an earthquake. Inertial sensors used by EEW algorithms provide unreliable measurement of very-low-frequency displacements (*Boore and Bommer,* 2005; *Melgar et al.,* 2013). *Meier et al.* (2016) found that the first few seconds of the P wave, as recorded by inertial sensors, do not contain enough information to forecast growth of the earthquake into a very large M8+ event. For example, during the 2011 $M_w$9.0 Tohoku-Oki, Japan earthquake, first and final alerts issued 8.6 s and 116.8 s after origin time underestimated the final magnitude by 1.8 and 0.9 magnitude units, respectively (*Hoshiba et al.,* 2011). This saturation resulted in underestimated ground motions in the greater Tokyo area and timely, but severely underestimated, tsunami warnings - including amplitudes and geographic extent (*Hoshiba and Ozaki*, 2014).

Blewitt et al. (2006) first proposed using geodetic measurements to overcome magnitude saturation after severe underestimation of the 2004 $M_w$9.3 Sumatra earthquake by long-period seismic observations within the first hour after the event. Based on this principle, instead of traditional seismic data (velocity and acceleration), geodetic EEW algorithms use observations collected by Global Navigation Satellite Systems (GNSS). GNSS can be conceptualized as strong-motion displacement sensors capable of measurement at the longest periods down to the static or permanent offset at 0 Hz (*Melgar et al.,* 2013). Since the 2004 Sumatra earthquake and tsunami, many GNSS-based techniques have been developed and improved to estimate source properties for earthquake and tsunami early warning in real-time (e.g., Allen & Ziv, 2011; Grapenthin & Freymueller, 2011; Colombelli et al., 2013; Grapenthin et al., 2014a,b; Minson et al., 2014; Crowell et al., 2016; Kawamoto et al., 2017); a history of geodetic early warning methods and their development can be found in *Bock and Melgar* (2016).

Seismic systems, as compared to geodetic, are also more limited by network configuration as they require good azimuthal coverage and dense station spacing. Out-of-network and edge-of-network events, i.e., those with poor azimuthal coverage, are often severely mischaracterized both in location and magnitude. This limitation is demonstrated by the performance of seismic point-source (e.g., ElarmS, *Allen et al.,* 2009b) and seismic finite-fault algorithms (e.g., FinDer, *Bose et al.,* 2012) during replays of the out-of-network $M_w$7.2 El Mayor-Cucapah occurring south of the US-Mexico border. When replaying the earthquake using only the stations operating in real-time within the US in 2010, both ElarmS and FinDer resulted in severe event mislocation and magnitude underestimation (*Ruhl et al.,* 2017). Geodetic systems, however, successfully characterized the El Mayor-Cucapah event using a similar (US-only) network geometry (Allen & Ziv, 2011; *Grapenthin et al.,* 2014a; *Ruhl et al.,* 2017). *Chung et al. (2017)* presented EEW results for recent earthquakes using two versions of ElarmS (E2 and E3). That work shows that while the number of missed and false events is substantially reduced in E3, both versions demonstrate that missed and false events are primarily those which originate outside of network boundaries (e.g., offshore) or in sparse network areas (e.g., eastern Oregon and Washington). Similarly, during replays of a simulated $M_w$8.7 megathrust earthquake on the offshore Cascadia subduction zone, ElarmS first locates the event offshore with an initial magnitude of ~8 before relocating it to within the network and lowering the magnitude to ~7. Geodetic finite-fault results for this simulation as well as the El Mayor-Cucapah event are more robust in this regard and demonstrate its ability to accurately estimate magnitudes of offshore, out-of-network events based on the first alerts produced from the seismic algorithm (*Ruhl et al.,* 2017). It is worth noting that in the western US several M>7 earthquake hazards exist near the edges of and beyond the footprint of the real-time EEW seismic network; for example, in the Cascadia subduction zone to the west, on crustal faults in the Basin and Range and Walker Lane to the east, and to the south and north of the contiguous US. To provide a true west-coast-wide system, we must extend the seismic network beyond the footprint of interest or include geodetic data to ameliorate some of the issues with limited network configurations.

To provide coverage for the full range of damaging earthquakes (M6+) in the US, several groups have developed EEW algorithms that make use of GNSS data. Three of these are currently being tested for implementation into ShakeAlert: G-larmS, BEFORES, and G-FAST (Grapenthin et al., 2014a; Minson et al., 2014; Crowell et al., 2016). In this work, we will focus on performance of the G-larmS algorithm (see *Murray et al., 2018* for a comparison) and make the data freely available so that other algorithm developers can conduct similar evaluations. The Geodetic Alarm System (G-larmS) was the first operational real-time geodetic system in the United States (*Grapenthin et al.,* 2014a), and has been running in real-time since the beginning of May 2014. G-larmS analyzes GNSS position time series in real-time, determines static offsets, and performs a least-squares inversion for slip using *a priori* fault geometries determined by the initial event location and magnitude provided by the ShakeAlert seismic algorithms (*Colombelli et al.,* 2013).

G-larmS operates as a triggered system and is coupled to the seismic point-source algorithms of ShakeAlert. Thus, our goal is to conduct end-to-end tests of the seismic only and the coupled seismic and geodetic systems with real data. We use a suite of large (M>6) earthquakes worldwide for which we have waveforms from both seismic and geodetic sites to test both the seismic (ElarmS) and geodetic systems (G-larmS). We quantify the timeliness and accuracy of seismic and geodetic magnitude and ground motion EEW alerts. We then show that the additional information and accuracy achieved by using available real-time GNSS data has substantial added value and that geodesy has an important role to play in providing warnings for the largest, most damaging earthquakes and their associated hazards.

## 2 Data

We test the EEW algorithms using 32 earthquakes from around the world ranging in magnitude from $M_w$6.0 (2004 Parkfield) to $M_w$9.0 (2011 Tohoku-oki) and with variable quantity and quality of seismic and geodetic data (Table 1). The database is dominated by subduction zone megathrust events but includes continental strike-slip (e.g., 2016 $M_w$7.0 Kumamoto), intraplate normal (e.g., 2017 $M_w$8.2 Tehuantepec), and other non-subduction zone events (e.g., 2015 $M_w$7.8

Nepal). The number of stations of each data type vary from a few to hundreds; both seismic and geodetic records (either real or synthetic) exist for 29 of the 32 earthquakes. We do not have seismic data for the 2010 $M_w7.7$ Mentawai earthquake, the 2014 $M_w7.7$ Iquique, Chile aftershock, or the 2015 $M_w7.3$ Nepal aftershock, for completeness we include these events as part of the geodetic analysis. The $M_w8.7$ "Cascadia001300" and $M_w7.0$ "Hayward4Hz" earthquakes are simulations of scenario events for which we have synthetic seismic and geodetic data (*Melgar et al.*, 2016, *Rodgers et al.*, 2018). For the $M_w6.9$ Nisqually 2001 event we have actual recorded seismic data as well as synthetic GNSS data from a slip inversion (*Crowell et al.*, 2016). In the following two sections, we discuss the details of seismic and geodetic data used in this study.

**Table 1.** List of earthquakes and their source parameters used in this study.

| | Event Name, Country | Origin Time (UTC)* | Longitude | Latitude | z (km) | Mw | Number of Geodetic Sites | Number of Seismic Sites | Mechanism |
|---|---|---|---|---|---|---|---|---|---|
| 1 | Tohoku2011, Japan | 2011-03-11T05:46:24 | 142.3720 | 38.2970 | 30.0 | 9.0 | 815 (288) | 211 | Reverse |
| 2 | Maule2010, Chile | 2010-02-27T06:34:14 | -72.7330 | -35.9090 | 35.0 | 8.8 | 27 | 7 | Reverse |
| 3 | Cascadia001300, USA (Synthetic) | 2016-09-07T07:00:00 | -124.6160 | 45.8638 | 19.8 | 8.7 | 62 | 40 | Reverse |
| 4 | Illapel2015, Chile | 2015-09-16T22:54:33 | -71.6540 | -31.5700 | 29.0 | 8.3 | 58 | 40 | Reverse |
| 5 | Tokachi2003, Japan | 2003-09-25T19:50:06 | 143.9040 | 41.7750 | 27.0 | 8.3 | 368 (189) | 313 | Reverse |
| 6 | Tehuantepec2017, Mexico | 2017-09-08T04:49:19 | -93.8990 | 15.0220 | 47.4 | 8.2 | 7 | 88 | Normal |
| 7 | Iquique2014, Chile | 2014-04-01T23:46:47 | -70.7690 | -19.6100 | 25.0 | 8.1 | 40 | 55 | Reverse |
| 8 | Ecuador2016, Ecuador | 2016-04-16T23:58:36 | -79.9220 | 0.3820 | 20.6 | 7.8 | 21 | 21 | Reverse |
| 9 | Kaikoura2016, New Zealand | 2016-11-13T11:02:56 | 173.0540 | -42.7370 | 15.0 | 7.8 | 39 | 34 | Strike-Slip |
| 10 | Nepal2015, Nepal | 2015-04-25T06:11:25 | 84.7310 | 28.2310 | 8.2 | 7.8 | 7 | 4 | Reverse |
| 11 | Ibaraki2011, Japan | 2011-03-11T06:15:34 | 141.2653 | 36.1083 | 43.2 | 7.7 | 1149 (432) | 278 | Reverse |
| 12 | Iquique_aftershock2014, Chile | 2014-04-03T02:43:13 | -70.4930 | -20.5710 | 22.4 | 7.7 | 17 | 0 | Reverse |
| 13 | Mentawai2010, Indonesia | 2010-10-25T14:42:22 | 100.1140 | -3.4840 | 20.0 | 7.7 | 13 | 0 | Reverse |
| 14 | N.Honshu2011, Japan | 2011-03-11T06:25:44 | 144.8940 | 37.8367 | 34.0 | 7.7 | 1148 (230) | 387 | Normal |
| 15 | Melinka2016, Chile | 2016-12-25T14:22:26 | -74.3910 | -43.5170 | 30.0 | 7.6 | 58 | 12 | Reverse |
| 16 | Nicoya2012, Costa Rica | 2012-09-05T14:42:08 | -85.3050 | 10.0860 | 40.0 | 7.6 | 9 | 14 | Reverse |
| 17 | Iwate2011, Japan | 2011-03-11T06:08:53 | 142.7815 | 39.8390 | 31.7 | 7.4 | 1149 (338) | 216 | Reverse |
| 18 | Miyagi2011A, Japan | 2011-03-09T02:45:12 | 143.2798 | 38.3285 | 8.3 | 7.3 | 892 (263) | 294 | Reverse |
| 19 | N.Honshu2012, Japan | 2012-12-07T08:18:20 | 144.3153 | 37.8158 | 46.0 | 7.3 | 978 (196) | 430 | Reverse |
| 20 | Nepal_aftershock2015, Nepal | 2015-05-12T07:05:19 | 86.0660 | 27.8090 | 15.0 | 7.3 | 5 | 0 | Reverse |
| 21 | ElMayor2010, Mexico | 2010-04-04T22:40:42 | -115.2800 | 32.2590 | 10.0 | 7.2 | 137 | 465 | Strike-Slip |
| 22 | Miyagi2011B, Japan | 2011-04-07T14:32:43 | 141.9237 | 38.2028 | 60.7 | 7.1 | 1137 (381) | 386 | Reverse |
| 23 | N.Honshu2013, Japan | 2013-10-25T17:10:18 | 144.5687 | 37.1963 | 56.0 | 7.1 | 59 (59) | 349 | Reverse |
| 24 | Puebla2017, Mexico | 2017-09-19T18:14:38 | -98.4890 | 18.5500 | 48.0 | 7.1 | 18 | 79 | Normal |
| 25 | Hayward4Hz, USA (Synthetic) | 2017-01-01T00:00:02 | -122.2850 | 37.9638 | 17.1 | 7.0 | 2301 (231) | 2301 | Strike-Slip |
| 26 | Kumamoto2016, Japan | 2016-04-15T16:25:05 | 130.7630 | 32.7545 | 12.5 | 7.0 | 277 (245) | 230 | Strike-Slip |
| 27 | Aegean2014, Greece | 2014-05-24T09:25:02 | 25.3890 | 40.2890 | 12.0 | 6.9 | 6 | 139 | Strike-Slip |
| 28 | Nisqually2001, USA (Synthetic Disp.) | 2001-02-28T18:54:32 | -122.7270 | 47.1490 | 51.8 | 6.9 | 26 | 63 | Normal |
| 29 | E.Fukushima2011, Japan | 2011-04-11T08:16:12 | 140.6727 | 36.9457 | 6.4 | 6.6 | 1146 (476) | 260 | Normal |
| 30 | Lefkada2015, Greece | 2015-11-17T07:10:07 | 20.6002 | 38.6650 | 10.7 | 6.5 | 23 | 4 | Strike-Slip |
| 31 | Napa2014, USA | 2014-08-24T10:20:44 | -122.3100 | 38.2150 | 11.0 | 6.1 | 224 (222) | 560 | Strike-Slip |
| 32 | Parkfield2004, USA | 2004-09-28T17:15:24 | -120.3700 | 35.8150 | 7.9 | 6.0 | 13 | 309 | Strike-Slip |

*Dates are formatted as year-month-day.

## 2.1 Seismic Data

Seismic data was collected from various sources for a total of 29 out of 32 earthquakes. The 2014 $M_w$7.7 Iquique, Chile aftershock and the 2015 $M_w$7.3 Nepal aftershock do not have seismic data and we were unable to include seismic data from the 2010 $M_w$7.7 Mentawai, Indonesia earthquake. We format the data into miniseed format in SI units (cm/s or cm/s/s) and create channel files specifying, among other things, the units, sample rates, and gains of each channel. Using the accompanying channel files, all waveforms for each event are combined and rewritten into one or more Earthworm tank-player files to be used for real-time replays.

For the 2001 $M_w$6.9 Nisqually, 2004 $M_w$6.0 Parkfield, 2010 $M_w$7.2 El Mayor-Cucapah, and 2014 $M_w$6.1 Napa earthquakes we downloaded acceleration and velocity waveforms in miniseed format along with channel files directly from the official ShakeAlert test suite (*Cochran et al.*, 2017). For each of these, waveforms begin two minutes prior to origin time and are a total of seven minutes long. Sampling rates range from 40 to 200 sps, depending on the instrument type.

Two events have only synthetic seismic data. The $M_w$8.7 "Cascadia001300" event is a simulated megathrust earthquake on the Cascadia Subduction zone offshore of Oregon, Washington, and California (*Ruhl et al.,* 2017). Acceleration waveforms (50 sps) begin one minute prior to origin time and are a total of 8.66 minutes long. The $M_w$7.0 "Hayward4Hz" earthquake is a simulated strike-slip rupture initiating on the down-dip extent of the Hayward fault in Northern California (Rodgers et al., 2018). Velocity waveforms sampled at 40 sps and with frequencies up to 4 Hz were obtained from *Rodgers et al.* (2018); each begins 2.0 seconds before origin time and has a total duration of approximately 1.5 minutes.

The remaining events are downloaded or obtained from local earthquake authorities in each country of origin (see Acknowledgements). Waveform lengths and sampling rates vary on a station-by-station and network-by-network basis; some are triggered stations that begin after the P wave.

### 2.2 Geodetic Data

The geodetic dataset consists of high-rate GNSS observations for 29 real earthquakes worldwide from the open dataset of *Melgar and Ruhl (2018)*. The displacement waveforms were calculated in a uniform fashion using the precise point positioning approach of *Geng et al.* (2013). The overwhelming majority of the recordings are collected at 1 sps but a few (2010 $M_w$7.2 El Mayor-Cucapah, 2012 $M_w$7.6 Nicoya, 2014 $M_w$6.1 Napa, and 2015 $M_w$7.8 Nepal) have some 5 sps recordings. These data were resampled to 1 sps for use with G-larmS which currently processes data by the integer-epoch. The data were processed into six-hour per-channel text-files to mimic the real-time trackRT format previously used at the Berkeley Seismological Laboratory.

We also use synthetic displacement data for three additional earthquakes. The $M_w$7.0 "Hayward4Hz" displacement data were created by integrating the seismic data simulated in *Rodgers et al.* (2018) and described in Section 2.1. The "Cascadia001300" synthetic data were developed using a hybrid semi-stochastic approach developed by *Melgar et al.* (2016) and described in *Ruhl et al.* (2017). The Nisqually 2001 earthquake is a real event in Washington state that was recorded seismically, but displacements were simulated by *Crowell et al.* (2016). For these events, data from multiple stations are rewritten into one time-ordered horizontal and one vertical component text file per event.

## 3 Methods

First, we replay seismic data from each earthquake through the ElarmS EEW algorithm in simulated real-time to estimate event magnitudes, epicentral locations, and origin times. We then use the seismic first-alerts, as well as "perfect" alerts (i.e., true origin time and location), to trigger the Geodetic Alarm System (G-larmS) and generate distributed slip and magnitude evolution time series. Using those results, we predict shaking intensity (MMI) time series for each seismic station for each event to compare to the observations. Finally, we employ an MMI-threshold approach (*Meier,* 2017) to accurately characterize warning times on a per-station basis, thus enabling classification of true positive, true negative, false positive, and false negative alerts for each event. Below we discuss the details of ElarmS (Section 3.1), G-larmS (Section 3.2), and the MMI-threshold method used for classifying real-time alerts (Section 3.3).

### 3.1 Seismic Alerts: ElarmS

ShakeAlert's seismic point-source algorithm (EPIC) is a derivative of the Earthquake Alarm System (ElarmS), a network-based EEW algorithm developed at the Berkeley Seismological Laboratory (BSL) over the last 10 years (*Allen et al.,* 2009a*; Kuyuk et al.,* 2014). Because there are only minor functional differences between EPIC and ElarmS, we test our dataset using the latest version of ElarmS currently operating at the BSL (*Chung et al., 2017*). ElarmS identifies and associates triggers and locates events epicentrally assuming a fixed depth or set of depths (8 and 20 km used in this study). Next, the algorithm estimates event magnitudes based on P wave amplitudes and distances to its estimated epicenter. ElarmS then generates earthquake alerts when a minimum of 4 stations with at least 0.2 s of data meet region-specific spatial constraints (e.g., station density is taken into consideration).

We create earthquake tank-player files and channel files containing all data for each earthquake and run them through ElarmS in simulated real time. When ElarmS identifies an event, it outputs estimates of origin time, magnitude, and location, as well as solution information such as number of stations. As additional stations trigger and seismic data develop, ElarmS refines and adjusts its source parameters and issues updated alerts. We retain a list of alert parameters in a separate log file for each event.

### 3.2 Geodetic Alerts: G-larmS

The Geodetic Alarm System (G-larmS) incorporates real-time GNSS data into EEW systems (*Grapenthin et al.,* 2014a). In real time operation, G-larmS continuously analyzes positioning time series and is capable of ingesting both relative displacements (baselines) and absolute positions from precise-point-positioning (PPP) solutions. During an earthquake, the ShakeAlert seismic system issues event messages containing hypocenter and origin time that trigger G-larmS to estimate static offsets epoch-by-epoch at each site. Simultaneously, it inverts these static offsets for distributed slip on a finite-fault. The latest version of G-larmS builds a linear fault using region-specific *a priori* geometries and, in addition, attempts to fit the event by imposing slip onto nearby known faults, allowing for complex geometries. In the first case, G-larmS centers the model fault plane on the earthquake hypocenter provided by ShakeAlert and allows the fault to grow symmetrically based on scaling relationships (*Wells and Coppersmith,* 1994). Model fault plane orientations are predefined for expected tectonic regimes based on location and Green's functions are calculated in real time. This means that, for instance, for an event in the San Francisco Bay Area, it will be modeled using linear San Andreas fault (SAF) parallel, SAF conjugate, and SAF splay (+/- 5 degrees from SAF) geometries. For the latter case, so-called 'catalog faults' are built into the system by simplifying models of large faults (e.g., UCERF3, Field et al., 2014; Slab1.0, Hayes et al, 2012). Therefore, for the San Francisco Bay Area example, an event in Oakland, CA is modeled with slip imposed onto the San Andreas and Hayward faults (separately) as well as on the growing, linear regional geometries. One benefit of using catalog faults is that Green's functions can be pre-computed for fixed station sets saving computation time during inversion. Another benefit is that curving faults such as megathrusts or complex strike-slip faults (e.g., big bend of San Andreas fault) can be modeled more accurately than with the linear tectonic regime faults. At each epoch, the geometry that minimizes the model misfit to the data is selected as the preferred solution. A detailed description of the original algorithm can be found in *Grapenthin et al.* (2014a, b) and previous performance with synthetics offshore Cascadia can be found in *Ruhl et al.* (2017).

In both replay (real events) and simulation mode (synthetic events), G-larmS is run in two separate steps, rather than simultaneously estimating offsets and finite-fault parameters as in the real-time system. The first module is the Offset Estimator (OE), which calculates and stores the co-seismic (static) offsets, and the second is the Parameter Estimator (PE) that actually inverts the offsets for slip on a finite fault. This separation is more efficient in an offline, personal-computer-based implementation. We run both the OE and PE twice: once using ShakeAlert XML messages created from the seismic first alert to trigger G-larmS (described in Section 3.2.1), and then again using messages containing the exact hypocenter and origin time as a "perfect" alert to trigger G-larmS (Section 3.2.2).

The G-larmS OE uses the ShakeAlert style XML event message to determine a start time for offset estimation at each station within a specific radius based on the event location, magnitude, origin time, and a configurable wave speed. Because static offsets typically arrive with the S wave, choosing a shear-wave velocity (~3 km/s) is often the preferred or recommended approach. However, a comparison of finite-fault solutions based on offset estimations started at estimated P wave and S wave arrival, respectively, showed that starting the offset estimation earlier resulted in damping of early offsets and, therefore, damping of the finite-fault solutions (*Ruhl et al.,* 2017). This is acceptable since near- and intermediate-field oscillating dynamic displacements can sometimes inflate initial static displacement (i.e., offset) estimates. Also, using a faster velocity can account for error in origin times and locations (i.e., prevent missing initial offsets) and may be more representative of average crustal velocities for deeper events. In this paper, we use a velocity of 5.2 km/s for all events, regardless of location or tectonic setting. G-larmS then calculates and stores the mean displacement amplitude before the calculated start time. These are subtracted from average displacements following the start time to estimate static offsets.

For the real earthquakes, we reformat the data into six-hour GPS time series for each station-component and store them in text files in GPS time units (i.e., without leap second adjustments). These are ingested in batches by the OE in faster-than-real-time replays and offsets are written to additional log files to be ingested later by the PE. For the simulated displacement data sets, we store horizontal displacement data for all stations in one time-ordered text file and vertical displacements in another. Random noise ($\pm 2.5$ cm and $\pm 4.0$ cm for horizontal and vertical components, respectively) is added to the displacements as they are read and offsets are estimated and written into individual log files. Once all offsets are calculated, the PE reads the offset logs and begins the slip inversion in real time based on the ShakeAlert XML message and the region-specific fault configurations.

G-larmS calculates earthquake magnitudes at each epoch based on the overall fault geometry and amount of slip imposed on it. Outputs include subfault geometries and the magnitudes of strike-slip and dip-slip components of slip per subfault. We simplify this information by calculating a surface-projected perimeter around the subfault patches that have slip greater than 40% of the maximum slip; we ensure this perimeter includes the hypocenter, even if it is located on a subfault that has less than 40% of the maximum subfault slip amount. This is used to calculate distances to the fault necessary for GM prediction. We then calculate the dominant rake of the overall fault based on the amount of strike-slip and dip-slip and characterize the solution as a rectangular fault with pure and uniform reverse (90), normal (-90), dextral (-180), or sinistral (180) slip. This information is used for GM prediction (see Section 3.3).

### 3.2.1 ElarmS-Triggered G-larmS

For a true real-time comparison, we first trigger G-larmS using the ElarmS first alerts. These are referred throughout the text and in figures as "ElarmS-Triggered G-larmS" solutions. We use the magnitude, epicentral location, and origin time of the first ElarmS solution to build the XML event messages for each event; depths are fixed to either 8 or 20 km, depending on the first alert. G-larmS replays always begin at the origin time, therefore solutions are calculated for each epoch as soon as the estimated P-wave reaches the closest GNSS station. Because geodetic sites may be closer than seismic sites, this may result in unrealistically timed solutions (i.e., before the ElarmS solution exists). Therefore, we remove all solutions before the ElarmS first alert plus one epoch.

### 3.2.2 Perfectly-Triggered G-larmS

In addition, we calculate "perfect" G-larmS solutions assuming that we know exactly where each earthquake occurs at exactly the origin time. These are referred throughout the text and figures as "Perfectly-Triggered G-larmS" solutions. We use M6.0 for the initial magnitude of each event as well as the exact depth as reported in the published catalog locations (Table 1). We employ this approach to assess how much the simulated real-time environment degrades a "perfect" solution.

### 3.3 Real-Time Alert Classification: MMI-Threshold Approach

Earthquake early warning is inherently a ground-motion warning system because users' actions depend on the level of shaking intensity expected at their individual sites, rather than on the magnitude or location of the earthquake. Because most algorithms provide the latter information without ground-motion estimates, it is difficult to assess an EEW system based on source parameters alone. Instead, *Meier* (2017) suggested to develop quantitative metrics, such as warning time, and to classify alerts using shaking intensity thresholds on a per-station basis. We follow that approach and assess our results with respect to ground motion in addition to comparing magnitude estimates.

First, we process the data by converting each seismic waveform to instrumental Modified Mercalli Intensity (MMI) time series using the method of *Worden et al.* (2012) and combine the three ground motion directions to obtain one maximum-MMI envelope for each station. Next, for the predictions, we compute peak ground velocity (PGV) and peak ground acceleration (PGA) for each station from the ElarmS, ElarmS-Triggered G-larmS, and Perfectly-Triggered G-larmS solution time series at each epoch. For all three solution types and for all global earthquakes in our study, we use the ground motion prediction equations of *Abrahamson et al.* (ASK14, 2014). For both G-larmS and ElarmS we use the same site-specific $V_S30$ value extracted from a slope-based global database from the USGS (*Wald & Allen*, 2007). We use dip and rake angles simplified from the G-larmS finite-fault solutions based on the average rake of the subfaults. The dips are 15°, 90°, 90°, and 60° for any fault with primarily reverse dip-slip, normal dip-slip, sinistral strike-slip, and dextral strike-slip, respectively. The simplified rakes are 90°, 180°, -180°, and -90° for each fault type, respectively. The GM estimates are controlled by varying only three input parameters: each algorithm uses its own $M_w$ estimate per-epoch, site-specific distance metric $R_{JB}$, and fault width W. $R_{JB}$ is the closest horizontal distance to the surface projection of the rupture (G-larmS) or point-source location (ElarmS). For the two G-larmS solutions, we calculate $R_{JB}$ using the fault perimeter described in Section 3.2. We then calculate width W for all solutions from the pure rakes and $M_w$ estimates using empirical relationships from *Wells and Coppersmith* (1994). All other parameters are held constant for GM predictions calculated for all three algorithms.

We calculate and store PGA and PGV values as time series based on the solutions that update each epoch. PGA and PGV are converted and combined into a maximum-MMI envelope time series in the same manner as the observations. We compare the observed and predicted MMI time series with respect to a specific MMI threshold on a per-station basis at all the available seismic sites for any given event. If both the observations and predictions exceed the specified MMI threshold, the *warning time* (WT) is defined as the time difference between the observed threshold crossing and the time at which the threshold crossing was first predicted. For instance, long warning times indicate that the prediction came well before the site experienced shaking equivalent to the MMI threshold. A short warning time, on the other hand, means that MMI threshold-exceeding shaking follows quickly after its prediction. If the WT is negative, the threshold-crossing ground motion alert was issued after actual ground motion already exceeded the threshold. All of these cases are classified as true positives (TP), even if late. We calculate median WTs for all TP sites. If neither the observations nor predictions cross the threshold, no WT is calculated and it is classified as a true negative (TN) end-user experience. If the observation crosses the threshold, but the prediction does not, no WT is calculated and it is classified as a false negative (FN). And finally, if the prediction crosses the threshold, but the observations do not, no WT is calculated and it is classified as a false positive (FP). Classifying alerts into these four categories allows quantification of the performance of an EEW system in the GM space (Meier, 2017). We repeat the MMI-threshold calculations for all seismic station sites for all events using the ElarmS, ElarmS-Triggered G-larmS, and Perfectly-Triggered G-larmS solutions for thresholds from MMI 3 to 7.

We do not include "Hayward4Hz" in the MMI analysis because it has >2000 stations only at very close distances and would dominate the results. The "Cascadia001300" data, on the other hand, includes only 40 seismic sites over a range of distances and their influence on statistics is therefore representative of real observations.

## 4 Results

ElarmS results were obtained for 26 out of the 29 earthquakes with seismic data. Thus, we have a total of 26 first-alert triggers from which we recovered ElarmS-Triggered G-larmS solutions. The 2016 $M_w$7.8 Ecuador, 2015 $M_w$7.8 Nepal, and 2011 $M_w$7.7 N. Honshu, Japan earthquakes did not produce seismic alerts due to poor station coverage or very far distances to the few closest stations. We were able to compute Perfectly-Triggered G-larmS results for all 32 earthquakes. In the following sections, we first discuss the timeliness and accuracy of magnitude estimates from the three algorithms, then present the MMI-threshold results and discuss the timeliness and accuracy of the real-time alerts. Throughout this section, we use the 2015 $M_w$6.5 Lefkada, 2017 Mw7.1 Puebla, 2012 Mw7.6 Nicoya, 2011 Mw7.7 Ibaraki, 2015 M8.3 Illapel, and 2011 $M_w$9.0 Tohoku-oki earthquakes as examples to illustrate the performance of the various algorithms over a range of magnitudes. In addition to spanning the magnitude range that we tested, these events are recorded on as few as four and up to hundreds of stations and also exhibit different focal mechanisms: Lefkada is a continental strike-slip earthquake, Puebla is a relatively deep intra-slab normal earthquake, and the remaining four are reverse events. Individual results for these six events are shown in Figures 1, 5, and 6; results for all other events are available in the Supporting Information (Table S1, Figures S1-S32). Figures 2, 3, 4, and 7 compile results for all events from which statistics are computed.

### *4.1 Accuracy and Timeliness of Magnitude Estimates*

Magnitude time series plotted for ElarmS (blue), ElarmS-Triggered G-larmS (red), and Perfectly-Triggered G-larmS (magenta) solutions in Figure 1 demonstrate the algorithm performance for the six example earthquakes. Geodetic magnitude estimates tend to approach the final magnitude (green), tracking modeled magnitude evolutions derived from published moment rate functions (black, Figure 1). The difference between the blue ElarmS magnitude estimates and the red and magenta G-larmS solutions tends to increase with increasing magnitude (top-left to bottom-right in Figure 1), revealing saturation in the seismic-only solutions. Magnitude-binned and averaged errors also show a significant increase in ElarmS magnitude saturation beyond M7.5, while magnitude errors for G-larmS are more stable with respect to increasing magnitude (Figure 2).

Figure 3 demonstrates the significant magnitude accuracy improvement by comparing estimates at three stages: the first time an alert is available (i.e., the first alert), the alert at 30 seconds after origin time (if available), and the final alert in the replay (Figure 1, Table S1). Alert times are calculated relative to origin time. For ElarmS, the final alert time is the last update produced shortly after the last triggered station arrives. For both G-larmS runs, all solutions were estimated until 180 seconds after origin time, which is the final solution time. ElarmS magnitude errors are -1.0±1.0 for the first alert, -0.71±0.75 at 30 s, and -0.50±0.83 at the final update. ElarmS-Triggered G-larmS magnitude errors are -0.62±0.86 for the first alert, -0.26±0.73 at 30 s, and -0.14±0.65 at the final update around 180 s. Perfectly-Triggered G-larmS magnitude errors are -1.3±0.78 at the first alert, -0.40±0.80 at 30 s, and -0.001±0.33 at the final update at 180 s. The better accuracy of both ElarmS-Triggered and Perfectly-Triggered G-larmS results are reflected by the clear 1:1 ratio in the two bottom right panels of Figure 3.

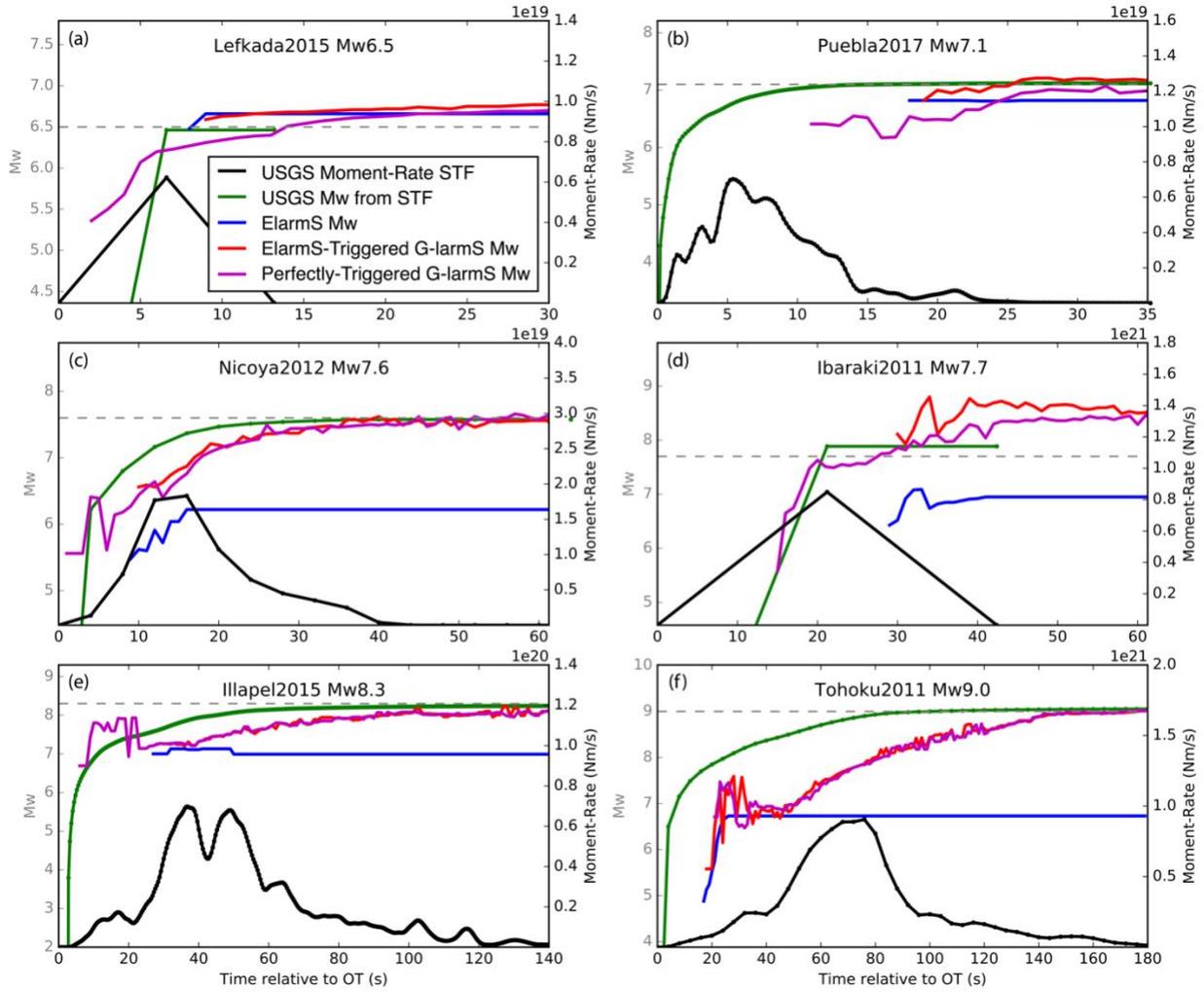

**Figure 1.** Magnitude estimate evolutions for six earthquakes used as examples throughout this paper. The magnitude of the examples increases from top left (2015 $M_w$6.5 Lefkada earthquake) to the bottom right (2011 $M_w$9.0 Tohoku-Oki earthquake). In each panel, the black curve is the USGS Moment-rate function derived from finite-fault source inversion or a triangle set to the width of twice the half-duration from moment tensors solutions published by the USGS. Green curves are the corresponding moment magnitude evolution for the black STF curves. Blue curves are from ElarmS, red curves are from ElarmS-triggered G-larmS solutions, and magenta curves are for Perfectly-Triggered G-larmS solutions. The dashed gray line shows the final Mw.

**Figure 2.** Mean magnitude error evolutions for six magnitude bins (a) $6.0 \leq M < 6.5$, (b) $6.5 \leq M < 7.0$, (c) $7.0 \leq M < 7.5$, (d) $7.5 \leq M < 8.0$, (e) $8.0 \leq M < 8.5$, (f) $8.5 \leq M \leq 9.0$. Errors are calculated as the predictions minus the observations such that a negative number shows magnitude saturation. Solutions were averaged for all events within the magnitude bin (total labeled) for ElarmS (blue) and ElarmS-Triggered G-larmS (red) solutions as the alerts came in. Points on each curve show

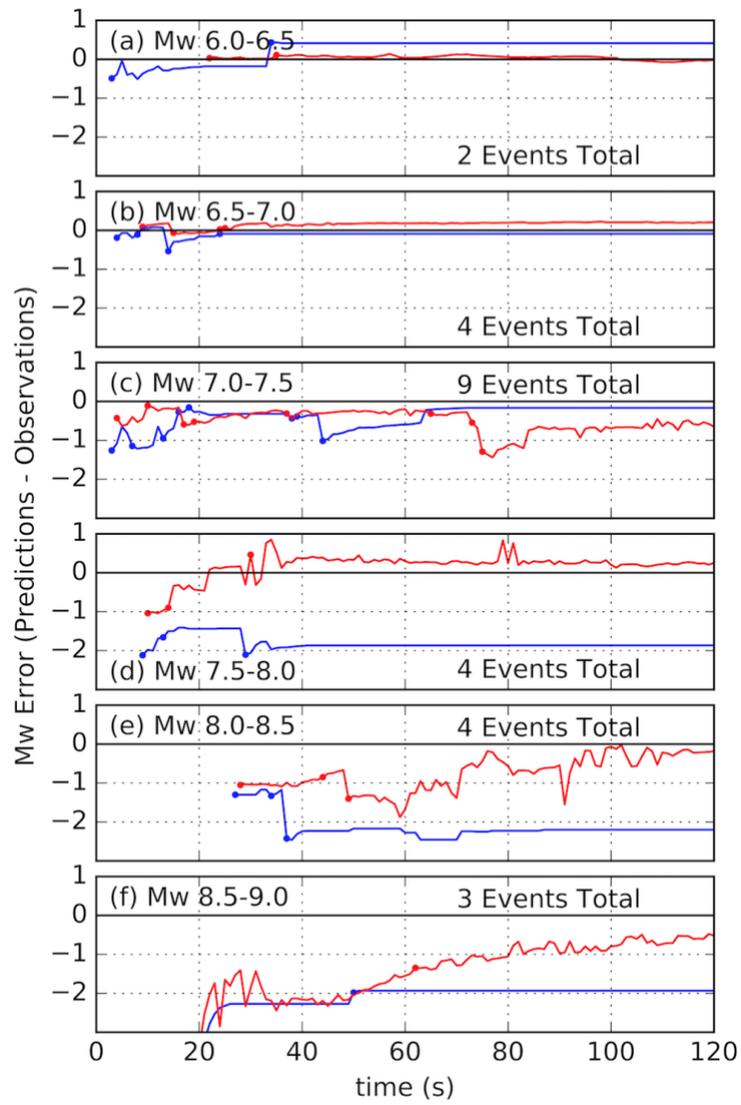

the first alert times of additional events included in each mean.

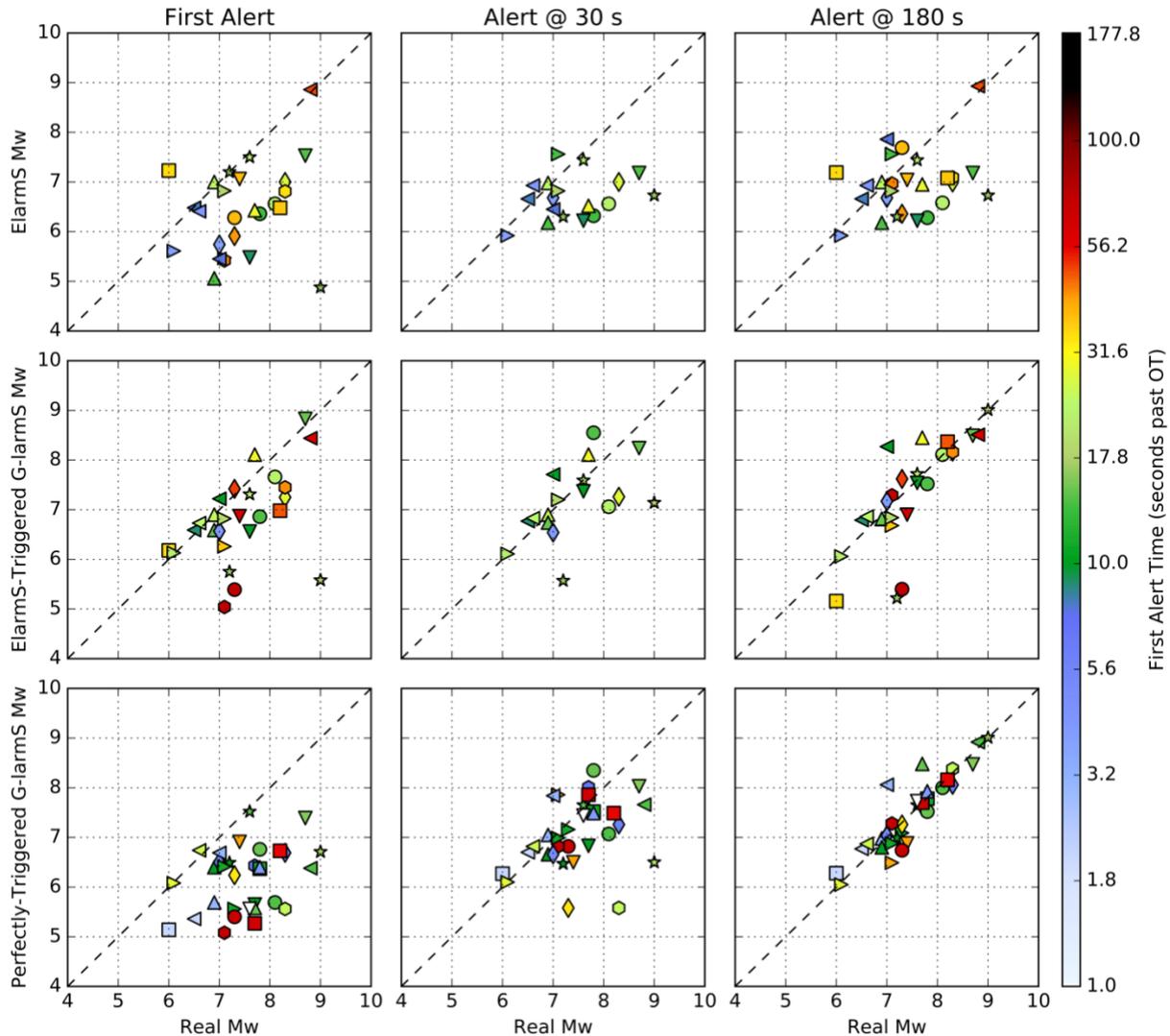

**Figure 3.** Magnitude estimates compared to real magnitudes for each algorithm for the first alerts (left column), alert at 30 s (middle column, where available), and at 180 s (or the final alert, right column). Top row subplots show ElarmS magnitude estimates, middle panels show ElarmS-triggered G-larmS magnitude estimates, and bottom panels show Perfectly-triggered G-larmS magnitude estimates. Dashed lines show 1:1 ratio in all subplots. Events are colored by the First Alert time in all three columns and maintain the same shape.

G-larmS provides an improvement of ~0.5 magnitude units, on average, by 30 s after origin time for all events (Figure 3). This improves with increasing magnitude as shown in Figure 2 and is often present in first alerts far earlier than 30 s. G-larmS first- and final-alert magnitude estimates are both statistically more accurate than ElarmS magnitude estimates, but as expected, the triggered geodetic algorithm takes longer, on average, to issue its first alerts than the seismic system (Figure 4a). Mean first alert times for ElarmS and ElarmS-Triggered G-larmS are 22 s ± 13.7 s and 31 s ± 20.5 s, respectively (Table S1 and Figure 4). Figure 4a does show that short first-alert times are achievable for the triggered geodetic system and that the bulk of the distribution indicates comparable first alert times for the respective algorithms.

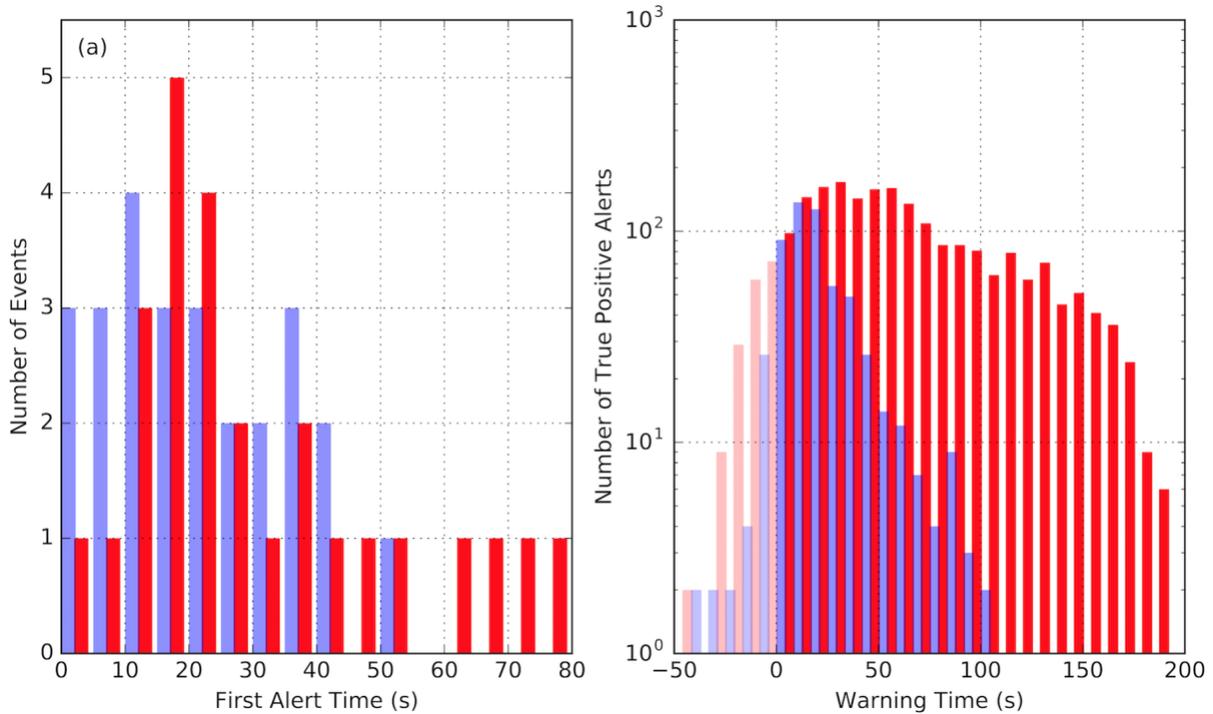

**Figure 4.** Histograms of (a) first alert times and (b) true positive warning times for ElarmS (blue) and ElarmS-triggered G-larmS (red) results with an MMI 4 threshold.

### 4.2 Accuracy and Timeliness of Ground Motion Estimates

In order to understand, from the end-user perspective, whether the improved, but delayed, geodetic earthquake characterizations are useful, we study the MMI performance of each solution. To demonstrate the technique, we show the MMI envelopes for two close-in stations per example earthquake with distances noted in Figure 5. The multi-colored curve with a black center line is the MMI envelope of each station and the gray shaded area shows the WT for the end-to-end test using a threshold of MMI 4. The shaded area spans from the time that either ElarmS or ElarmS-Triggered G-larmS MMI predictions exceed the MMI-threshold to the time when the data exceeds the same threshold. Sometimes the final shaking is overestimated (Figure 5a) and other times it is underestimated (Figure 5b), but the threshold approach enables us to look at accuracy in terms of binary alert classification. Even though ground motions may be over- or underestimated, as long as they are above or below the user's threshold of interest, the alert is useful and considered a success or true alert (made up of TPs and TNs). Geodetically inferred TP alerts are correctly issued for all 12 site examples in Figure 5; ElarmS, however, did not predict ground shaking stronger than the threshold at the bottom-right five stations. This means that, at least at these locations, it never issues a warning for users who will experience shaking greater than MMI 4.

Alert classifications for all stations recording the six example events are shown in Figure 6 with an MMI 4 threshold. For the smallest event, the 2015 $M_w$6.5 Lefkada earthquake, there is no significant improvement from the coupled seismic-geodetic solution (Figure 6a). This is not surprising since seismic saturation is not an issue at this magnitude, however, for the larger events, there is remarkable improvement in the ground motion estimates.

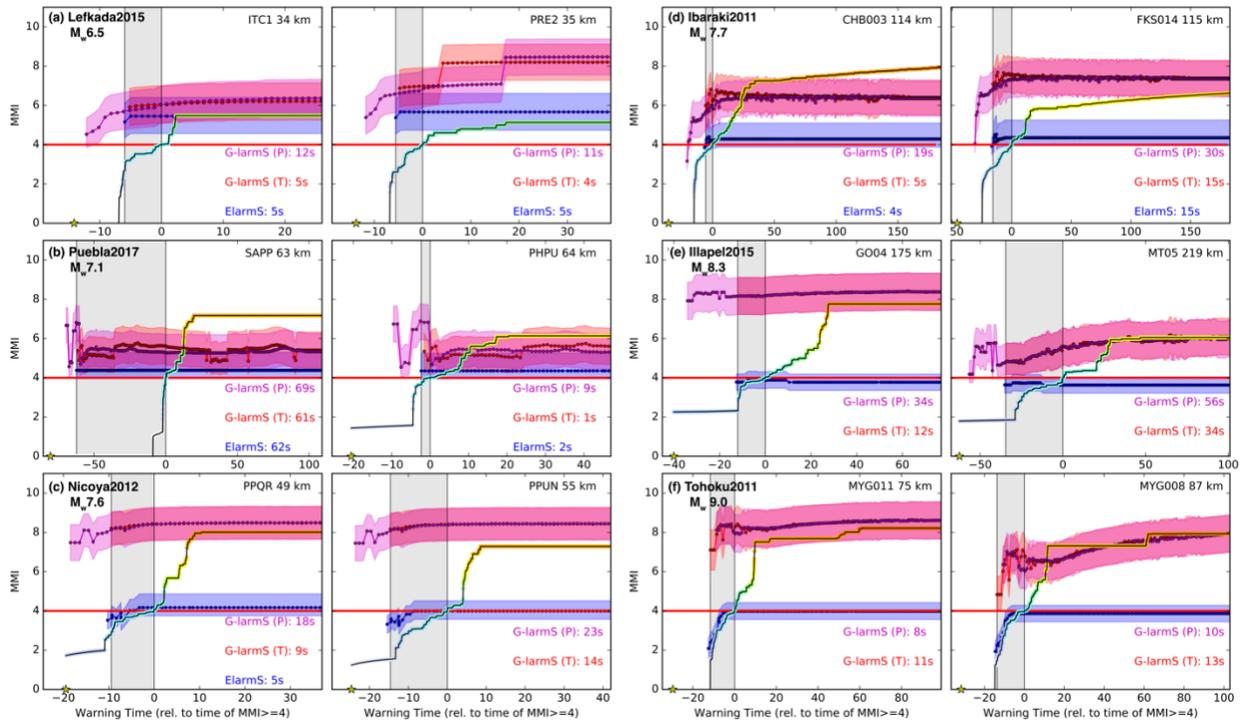

**Figure 5.** Examples of the MMI threshold method of alert classification for individual stations. For each of the six example earthquakes shown in Figure 1, we show two example stations. The station name and hypocentral distance are labeled in the top right corner of each figure. Red horizontal line shows the threshold of MMI 4. Colored curve shows the maximum MMI envelope of three component seismic data for each station. Blue curves with shaded regions show MMI estimates from ElarmS +/- 1 sigma. Red curves with shaded region show MMI estimates from ElarmS-triggered G-larmS solutions +/- 1 sigma. Magenta curves show MMI estimates from Perfectly-triggered G-larmS solutions +/- 1 sigma. Warning times calculated based on the crossing of the MMI threshold are shown in bottom right hand corner of each subplot for the three algorithms. The gray shaded regions show the maximum warning times achieved from either ElarmS or the ElarmS-triggered G-larmS solutions. Note that the Tohoku ElarmS-triggered G-larmS results arrive before the Perfectly-triggered G-larmS results; this is reflective of significant origin time error produced by the ElarmS first alert.

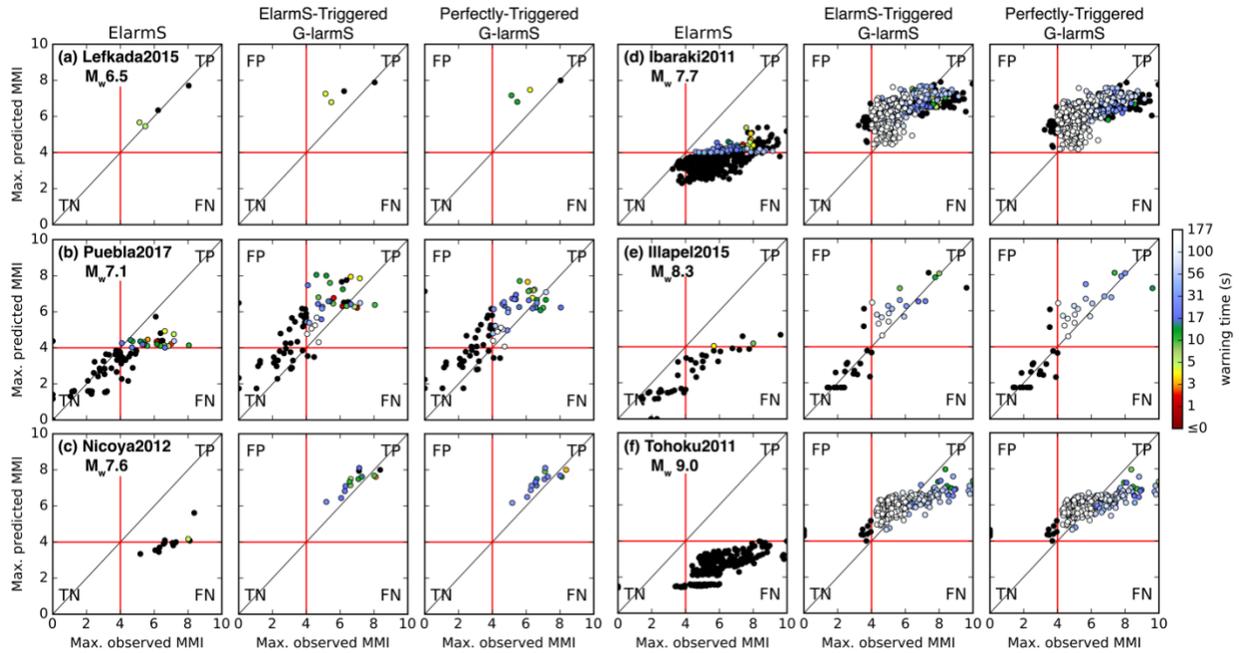

**Figure 6.** Real-time alert classification plots for the six example earthquakes shown in Figures 1 and 5. Each event has three subplots showing alert classifications for ElarmS, ElarmS-triggered G-larmS, and perfectly-triggered G-larmS solutions. Each data point is colored by the warning time at that station based on a threshold of MMI 4 (red lines). Quadrants are labeled as true positive (TP), true negative (TN), false positive (FP), or false negative (FN).

To further understand the performance of the algorithms, we synthesize the results from all events and all stations. As described in Section 3.3, we classify TP, TN, FP, and FN user-alerts and compute warning times (WT) for all TP sites for all events (Figure 7, Table 2). Using an alert threshold of MMI 4, ElarmS produced 12.3% TPs with a median WT of 16.3 s. Meanwhile, ElarmS-Triggered G-larmS resulted in 44.4% TPs, a 31% increase over ElarmS. Median WT increased to 50.2 s – more than twice the WT of ElarmS alone. The significant increase in warning times is demonstrated for MMI 4 in Figure 4b. As shown in Table 2, the total percentage of true alerts (%T=%TN+%TP) increases from 52.0% for ElarmS using MMI 4 to 71.8% when incorporating ElarmS-triggered G-larmS solutions; the number of false alerts decreases by 20% when using ElarmS-triggered G-larmS. ElarmS-Triggered G-larmS classification and timeliness results are very similar to, and nearly indistinguishable from, Perfectly-Triggered G-larmS results (Figures 6 and 7), even when the individual station estimates may be very different through time (e.g., Figure 7a).

**Table 2.** Table of real-time alert classification results from ElarmS (top bold row), ElarmS-Triggered G-larmS (middle bold row), and ElarmS-Triggered G-larmS (bottom bold row) using thresholds of MMI 3, 4, and 5 (left, middle, and right columns, respectively). The setup of each bold panel is shown in the top right panel. Percentages of true positive (TP), true negative (TN), false positive (FP), and false negative (FN) alerts are shown in the first four quadrants, the total percentage of true (%T) and false (%F) alerts are shown in the next row, and warning time (WT) medians and standard deviations (StdDev) are shown for all TP alerts. Finally, we show two cost savings performance metrics Q for cost-ratios r of two and ten.

| Setup: | | | MMI 3 | | MMI 4 | | MMI 5 | |
|---|---|---|---|---|---|---|---|---|
| % FP | % TP | | | | | | | |
| % TN | % FN | | | | | | | |
| % T | % F | | | | | | | |
| Median +/- StdDev | | | | | | | | |
| Q (r=2) | Q (r=10) | | | | | | | |
| **ElarmS** | | | 3.1 | 47.8 | 1.2 | 12.3 | 0.1 | 2.1 |
| | | | 10.9 | 38.1 | 39.7 | 46.8 | 70.6 | 27.2 |
| | | | 58.7 | 41.2 | 52.0 | 48.0 | 72.7 | 27.3 |
| | | | 31.2 +/- 30.3 | | 16.3 +/- 20.9 | | 6.4 +/- 13.5 | |
| | | | 0.04 | 0.50 | -0.62 | 0.12 | -0.86 | -0.03 |
| **ElarmS-Triggered G-larmS** | | | 7.0 | 78.0 | 13.5 | 44.4 | 18.5 | 23.2 |
| | | | 7.0 | 8.0 | 27.4 | 14.6 | 52.1 | 6.1 |
| | | | 85.0 | 15.0 | 71.8 | 28.1 | 75.3 | 24.6 |
| | | | 36.9 +/- 48.6 | | 50.2 +/- 49.9 | | 57.7 +/- 49.0 | |
| | | | 0.65 | 0.88 | 0.05 | 0.67 | -0.68 | 0.63 |
| **Perfectly-Triggered G-larmS** | | | 7.2 | 80.6 | 12.5 | 44.1 | 16.5 | 22.4 |
| | | | 6.9 | 5.4 | 28.4 | 14.9 | 54.1 | 6.9 |
| | | | 87.5 | 12.6 | 72.5 | 27.4 | 76.5 | 23.4 |
| | | | 43.7 +/- 46.9 | | 42.8 +/- 55.8 | | 49.3 +/- 54.3 | |
| | | | 0.71 | 0.91 | 0.07 | 0.67 | -0.60 | 0.61 |

Choosing MMI thresholds of 3 and 5 (Figure 7) also leads to 20-30% increases in TPs when using G-larmS in addition to the seismic system, although the total increase of true alerts (%T) is only increased by ~3% for MMI 5. This reflects a large increase in TPs coincident with a decrease in TNs at the higher MMI 5 threshold (Table 2). The percentage of FNs is decreased by the geodetic system, but the percentage of FPs increases. The percentage of TPs is also higher for both algorithms when using an MMI 3 threshold (47.8% for ElarmS compared to 78.0% for ElarmS-triggered G-larmS). Conversely, the percentage of TPs is lower for both algorithms when using an MMI 5 threshold (2.1% compared to 23.2%). For ElarmS, median warning times increase to 31.2 s using MMI 3 and decrease to 6.4 s using MMI 5. This shows that for a seismic system, a lower shaking intensity threshold will result in more TPs with longer WTs than a higher threshold with fewer TPs and shorter WTs. Notably, the ElarmS-Triggered G-larmS system results show an inverse relationship to MMI-threshold choice: median warning times decrease to 36.9 s when using MMI 3 and stay roughly the same when increasing from MMI 4 to 5 (50.2 s at MMI 4 compared to 57.7 s at MMI 5). Longer median warning times are achieved using perfect alerts rather than triggered alerts: perfectly-Triggered G-larmS produces TP WTs of 43.7 s at MMI 3, 42.8 s at MMI 4, and 49.3 at MMI 5.

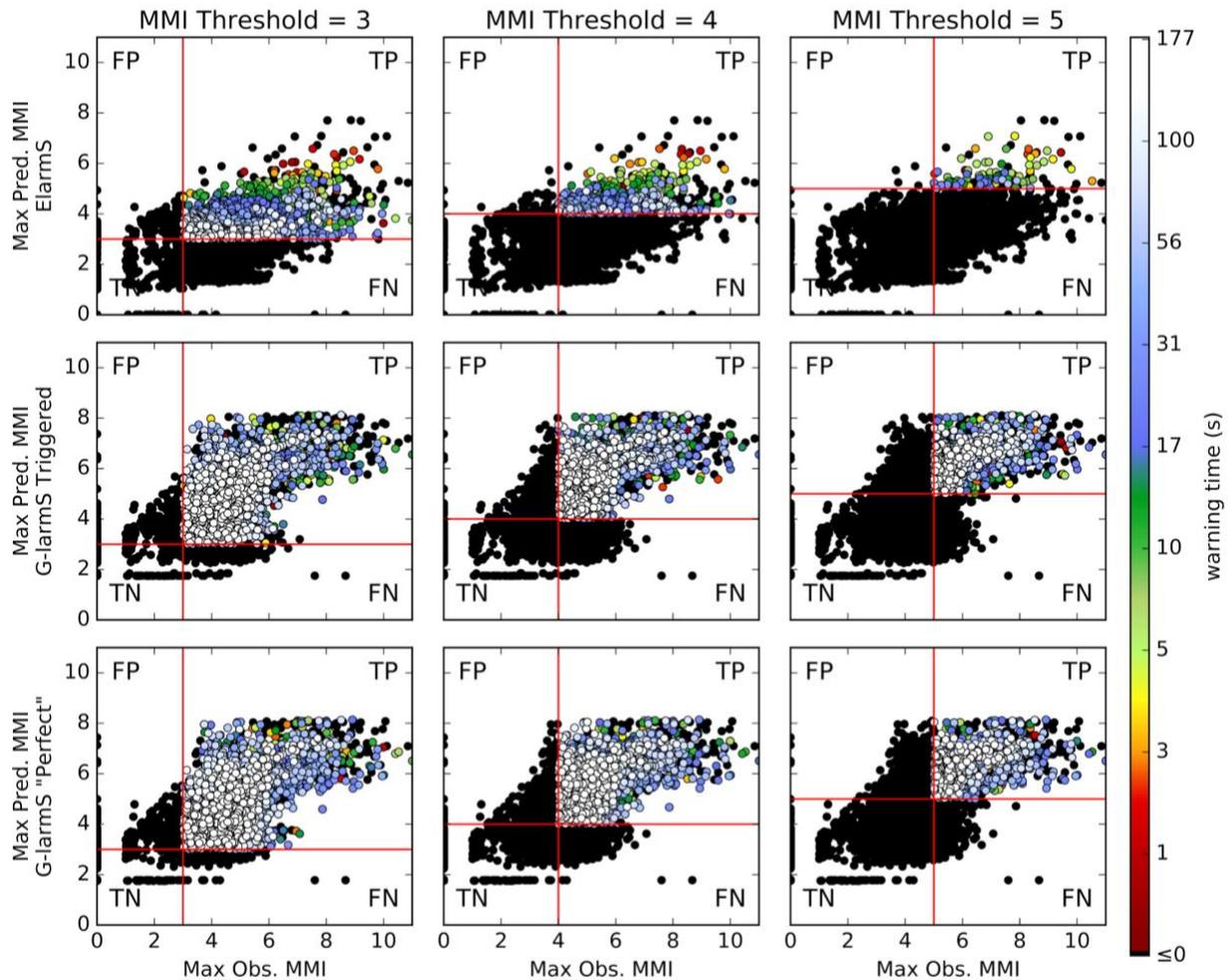

**Figure 7.** Real time classification plots for thresholds of MMI 3 (Left Column), MMI 4 (Middle Column), and MMI 5 (Right Column) for all 5151 individual station records for 31 earthquakes total ("Hayward4Hz" synthetic excluded, see text for discussion). Top row shows ElarmS results, middle row shows ElarmS-triggered G-larmS results, and bottom row shows perfectly-triggered G-larmS results. Data are colored by warning time based on the specific thresholds used (red lines). Quadrants are labeled as true positive (TP), true negative (TN), false positive (FP), or false negative (FN).

## 5 Discussion

As shown in Figure 7, when weighed against real data, the performance of a coupled seismic-geodetic system is better than a seismic-only system; useful warning times are routinely achievable for large events with high ground motions. Earthquake source processes can be complex and ground motion prediction equations have substantial uncertainties attached to them, thus some proportion of false alerts remain (Table 2); EEW systems and users that consider and effectively deal with these uncertainties based on some previously defined tolerance level will be important going forward. For example, a user's tolerance can be quantified by a cost ratio **r** defined as the ratio between the cost of damage to the cost of taking action to avoid said damage (Aagard et al., 2018). A value of r>>1 represents a user that is false alert tolerant, while r < 1 is representative of a user that is intolerant to false alerts (it costs more to take action than to sustain damage). Consider a nuclear power plant, it may be very costly to shut down, therefore, the cost of action is very high in comparison to the cost of potential damage from uncertain shaking estimates; this would result in a very small **r** value for that particular user. For students in a classroom, however, the cost of getting under a desk for a few seconds to minutes has a very small cost compared to the cost of injury or death that may result during very strong shaking; this scenario results in a very high cost ratio and a higher tolerance for false alerts. Likewise, slowing or stopping a train is a relatively low cost of action compared to the train potentially derailing as a result of strong ground shaking; there are many such scenarios where the cost ratio is much greater than one.

One way to quantify the success or usefulness of an EEW system is to calculate the Cost Savings Performance Metric Q for theoretical users with different cost ratios (Aagard et al., 2018). Q is defined as a function of the cost ratio and real-time alert classification percentages:

$$Q = \frac{\%TP - \frac{\%FP + \%F}{r-1}}{\%TP + \%FN} \qquad (1)$$

Note that this metric is user-dependent, as its value will change based on each user's cost ratio, r, and thus, when the respective Q is greater than zero, the EEW system is useful for that user and provides a cost savings. Using the real-time alert classifications for each algorithm shown in Table 2, we calculate Q using two cost ratios (r = 2 and r = 10) at the three MMI thresholds for each algorithm. For a cost ratio of two, ElarmS alerts are only slightly useful at MMI 3 (Q = 0.04) and not useful at the higher MMI thresholds (negative Q values). For a cost ratio of ten, seismic alerts result in positive Q values for an MMI 3 threshold, but decrease with increasing MMI to the point of no longer being useful at MMI 5 (Table 2, Figure 8). Geodetic solution metrics, on the other hand, are positive for all MMIs between 3 and 5 using a cost ratio of ten and are positive for a ratio of two at MMI thresholds of 3 and 4. The highest Q values are found for the geodetic algorithms at MMI 3 (0.88 for ElarmS-Triggered G-larmS and 0.91 for Perfectly-triggered G-larmS), suggesting that both systems will be more successful at alerting for larger shaking intensities accompanying large events when using a lower intensity alert threshold.

Additionally, using a large cost ratio (e.g., 10 or greater) representative of low-cost actions such as personal response (getting under a desk) or mass industrial response (stopping trains and elevators) shows that geodetic algorithm solutions are always useful, with a minimum Q value of 0.61 using a threshold of MMI 5. These numbers also show that geodetic solutions are much less affected by increasing the MMI threshold than ElarmS is for large events. We did not include smaller magnitude earthquakes (M<6) in our analysis, therefore these statistics may only apply to higher magnitude events with large ground motions.

Figure 8 shows the performance metric results for a large range of cost ratios for the same MMI thresholds. Most striking is that the coupled seismic-geodetic approach generally results in higher cost-savings at smaller cost ratios for similar MMI thresholds. This is particularly important as r approaches 1, emphasizing that geodetic analyses provide significant value to users with small false alert tolerances.

Future studies should include a Q analysis of the full range of magnitudes to make proper recommendations about the use of the ShakeAlert system. Of course, much work is also needed in order to quantify the actual values of the cost ratio, r, for a variety of users; for the present, we can only speculate as to what these might be. However, using the framework of Equation 1 it is clear that geodetic algorithms provide substantial added value to EEW for M>7 earthquakes.

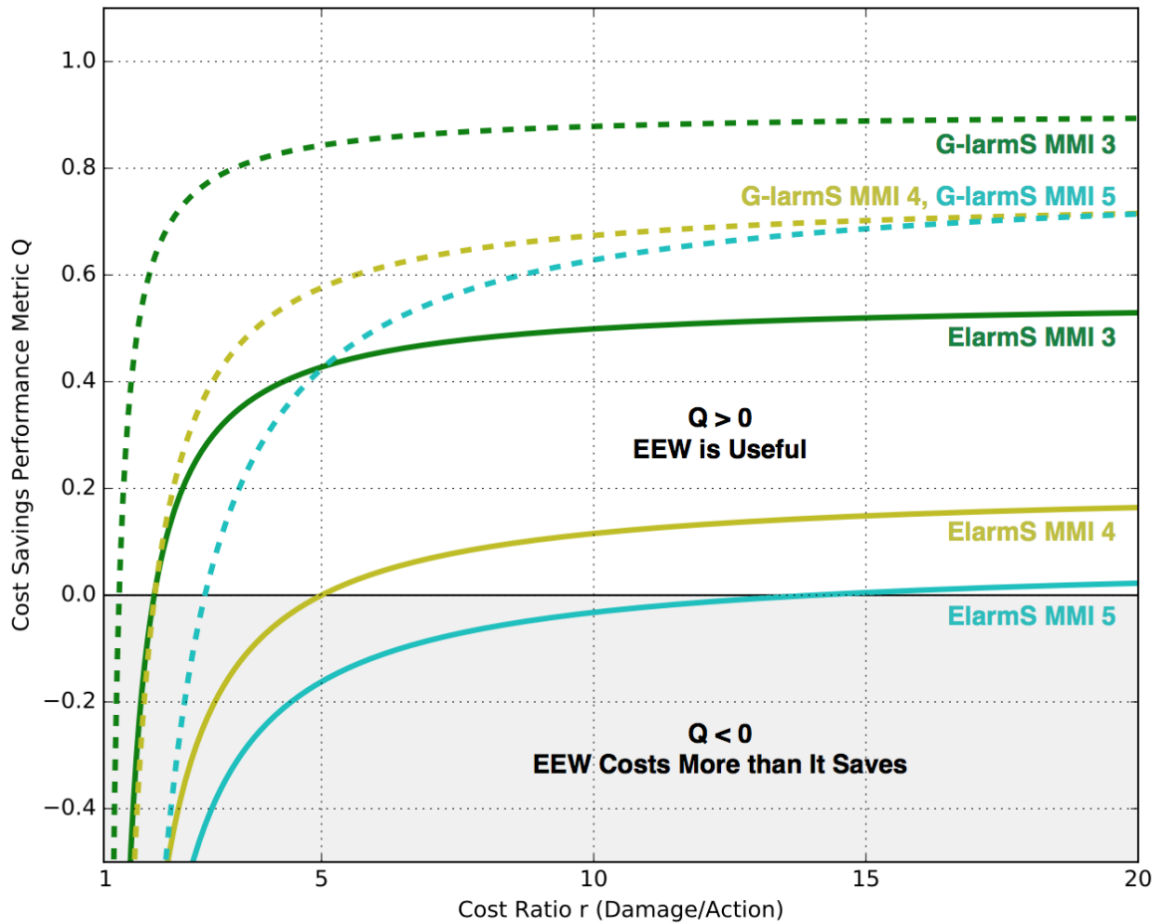

**Figure 8.** Cost Savings Performance Metric Q as a function of cost ratio r for ElarmS (solid) and ElarmS-Triggered G-larmS (dashed) real-time alert classification results using thresholds of MMI 3 (green), MMI 4 (yellow), and MMI 5 (cyan).

Geodetic algorithms provide more accurate GM estimates (Figure 7) and they also provide better user-alert timeliness (Figures 4b). If we consider the true alerts, the geodetic system provides improved warning times and accuracy under all thresholds tested against for a larger portion of sites than ElarmS does (Table 2, Figure 4b). Geodetic warning times can also be very long (>30 s), even for high MMI thresholds (Figures 3, 4b, 5, and 7).

Not only are the median warning times longer when incorporating geodetic finite-fault solutions, but, perhaps more importantly, for some of the stations with the largest ground motion, ElarmS never reaches the MMI threshold (e.g., Figure 5f). Thus, ElarmS never issues an alert for some of the strongest ground shaking. This behavior results in more missed alerts or false negatives than the geodetic algorithm and a worse overall performance for the seismic system, especially for users sensitive to false negatives.

Because they rely on the first few seconds of the P wave, initial alerts from seismic EEW algorithms are faster than those from geodetic algorithms (Figure 4a). The higher noise levels in real-time geodetic time series (1-3 cm) preclude detection of the early onset signals, and increase the delay until measured ground motion exceeds the noise. However, seismic P wave methods have important limitations. As shown in the previous section, they underestimate the magnitudes of very large earthquakes, and as a result of this and the point source assumption, are not as useful for providing accurate ground motion alerts to users for these events. As illustrated during the Mw9.0 Tohoku-oki earthquake, using seismic algorithms alone would mean that the region of predicted strong shaking is significantly smaller than the actual area of strong shaking and many users would not receive an alert. In an effort to ameliorate this, *Hoshiba & Aoki* (2015) proposed the PLUM method which uses a very dense network of strong motion sensors and uses present observations of the seismic wavefield to forecast its likely intensity some time into the future. This algorithm, now operational in Japan, improves EEW performance during large events but produces only very short, fixed warning times (~10 s) and is limited to very dense networks.

Thus, there is a need for faster unsaturated magnitude calculations that can be translated to ground motion estimates. GNSS fills this niche: though slower to create an initial alert, the geodetic EEW algorithms provide significantly more accurate magnitude estimation, slip distribution estimation, and, in turn, more accurate ground motion prediction for large earthquakes. A common refrain is that GNSS source estimates are too slow to be useful in EEW, our results show that while this is true for moderate magnitude events around ~M6.5, for the larger earthquakes, GNSS provides a substantial and very valuable improvement in the timeliness and reliability of the alerts. Geodetic algorithms can be used to correctly warn a significantly larger portion of the population than seismic algorithms alone, and to provide substantial warning times for users in areas that will experience strong shaking. Our results show that the current ShakeAlert point-source system alone will not be sufficient to forecast the strongest shaking due to the largest events. Instead, combining seismic and geodetic approaches can capture the earliest shaking with shorter warning times using the seismic data and predict more distant shaking with longer warning times from the geodetic data. We recommend a combined system as well as one that continually updates to account for the growth of large earthquakes. Large earthquakes are complex geophysical phenomena, and the best outcome for EEW is obtained when they are measured by geophysical instrumentation with complementary strengths. Our results conclusively show that GNSS add substantial value to seismic EEW systems.

Finally, we note that we have only produced testing results for one candidate geodetic finite-fault algorithm. As discussed in Section 2, there are other candidate algorithms and we hope that they can be tested in a similar fashion with the data we make available so that we can make objective comparisons between the proposed algorithms and determine which features of the respective methods produce the most reliable ground motion estimates with the longest warning times.

## 6 Conclusions

Here, we quantified the timeliness and accuracy of seismic and geodetic magnitude and ground motion EEW alerts by testing a suite of large (M>6) earthquakes worldwide. ElarmS magnitude errors indicate magnitude saturation for large events and are -1.0 ± 1.0 and -0.50 ± 0.83 units for the first and final alerts, respectively. ElarmS-triggered G-larmS magnitude errors are -$0.62 \pm 0.86$ and $-0.14 \pm 0.65$ units at the first and final update, respectively. We calculated shaking intensity time series for each station for each event using the simulated real-time solutions. Applying an MMI-threshold approach to accurately characterize warning times on a per-station basis, we classified true positive (TP), true negative, false positive, and false negative (FN) alerts for each event. Using a threshold of MMI 4, ElarmS produced only 12.3% TP alerts with a median warning time of $16.3 \pm 20.9$ s, while ElarmS-triggered G-larmS solutions result in 44.4% TP alerts with a longer median warning time of $50.2 \pm 49.8$ s. The number of missed alerts (FN) using thresholds of MMI 3 and 4 is reduced by over 30% with the seismically-triggered geodetic EEW system. Perfectly-triggered G-larmS results provided similar statistics as the triggered system, with slightly longer warning times and more accurate final magnitudes.

Analysis of the cost savings performance metric Q showed that the geodetic solutions provide a higher cost-savings value to users with a variety of cost ratios r when compared to the seismic-only system, particularly for less false-alert tolerant users. It also suggests that both systems will be more successful at alerting for larger shaking intensities accompanying large events when using a lower intensity alert threshold. These results demonstrate the added value of a geodetic EEW system, quantifying improvements in magnitude accuracy, ground motion accuracy, and alert timeliness in ground motion space. Permanent, static surface displacements are an essential part of earthquake observation and must be incorporated, having been measured with fidelity via GNSS, into EEW systems to ensure their success for the largest, most damaging earthquakes.

## Acknowledgements and Data

This work was funded by the Gordon and Betty Moore Foundation through Grant GALA # 3024 to U.C. Berkeley, and the U.S. Geological Survey Cooperative Agreement G17AC00346. We thank the network operators who processed and supplied the data to make this study possible. Acceleration data for the 11 Japanese earthquakes are available through the online portal of the K-NET and KiK-net strong-motion seismograph networks operated by the National Research Institute for Earth Science and Disaster Resilience, Japan. The Chilean data are from Centro Sismológico Nacional, Universidad de Chile. The two 2017 Mexican earthquakes were obtained from the National Seismological Survey (SSN) and Engineering Institute at the National Autonomous University of Mexico (UNAM). The 2014 $M_w$6.9 Aegean data were downloaded from the Strong Ground-Motion Database of Turkey. The 2015 $M_w$6.5 Lefkada earthquake data are from the National Observatory of Athens. The 2012 $M_w$7.6 Nicoya data are from the Observatorio Vulcanológico y Sismológico de Costa Rica (OVSICORI). The 2015 $M_w$7.8 Nepal data are from *Galetzka et al.* (2015). The 2016 $M_w$7.8 Ecuador earthquake is from Instituto Geofisico de Ecuador. Finally, the 2016 $M_w$7.8 Kaikoura data are from GNS Science in New Zealand. The geodetic data was collected from various sources and is published by Melgar and Ruhl (2018). The seismic data has been permanently stored at https://zenodo.org/record/1469833. There is one large tarball available for download. Inside the archive, there is one folder per event clearly labeled with the names in Table 1. Inside each event folder is a textfile (EVENT_sm.chan) with station metadata (station codes, coordinates, gains, sampling rates, etc.). There are "vel" and "accel" folders, where velocity and acceleration data are available, respectively. Each contains miniseed data named as STA.CHAN.mseed, where STA is the station code and CHAN is the channel code. Future studies that use our dataset should cite this paper.

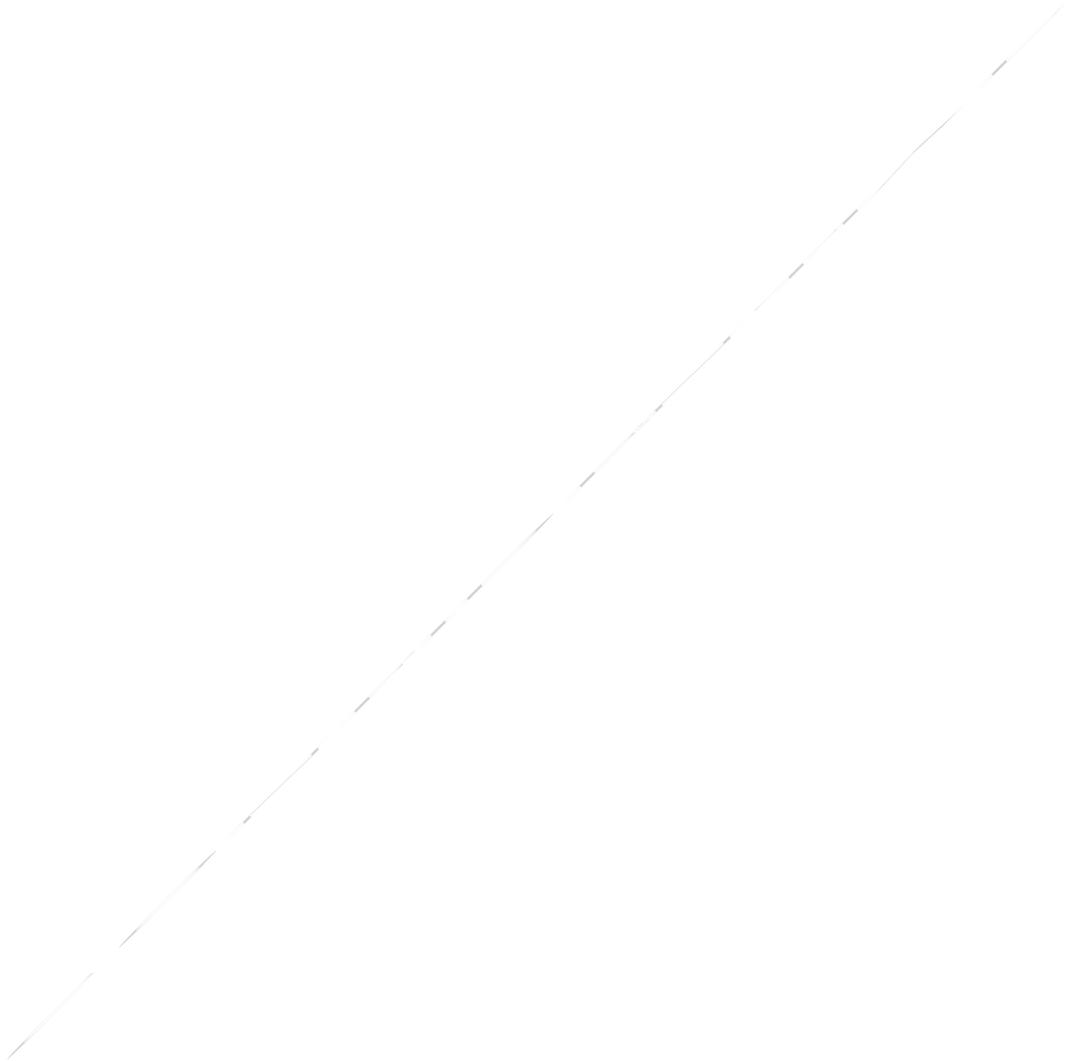